\documentclass[aps,prl,twocolumn,showpacs,superscriptaddress,groupedaddress]{revtex4-1}
\usepackage{graphicx}
\usepackage{rotating,psfrag}
\usepackage{amsfonts,amsmath,color}
\usepackage{amssymb}
\usepackage{tabularx}

\newcommand{\bx}{{\bf {x}}}

\newcommand{\rme}{\mathrm{e}}
\newcommand{\bu}{{\bf {u}}}

\newcommand{\rmi}{\mathrm{i}}
\newcommand{\Fra}[2]{\displaystyle \frac{ #1}{ #2}}

\newcommand{\Dpar}[2]{\displaystyle \Fra{\partial #1}{\partial #2}}

\newcommand{\Dparxy}[3]{\displaystyle \Fra{\partial^2 #1}{\partial #2 \partial #3}}

\newcommand{\Lapl}[1]{\displaystyle \boldsymbol{\nabla}^2 #1}

\newcommand{\Grad}[1]{\displaystyle \boldsymbol{\nabla} #1}

\newcommand{\Vmy}[1]{\mathbf{#1}}
\begin{document}

\title{Helicity transfer during quantised vortex reconnection}

\author{Andrew W. Baggaley}
\affiliation{
School of Mathematics and Statistics, University of Glasgow,
Glasgow, G12 8QW, UK
}

\date{\today}

\begin{abstract}
We consider the reconnection of two untwisted, linked quantised vortex rings in a Bose-Einstein condensate. We show that the reconnection is capable of transferring helicity, from the links present in the initial configuration, to a twisting of the resulting vortex ring, and hence a rotation of the phase plane along the vortex ring. As velocities in a quantum fluid are the gradient of the phase, a twisting of the phase along the vortices leads to an axial flow along the vortices. Hence one would expect that the dynamics of quantum turbulence are strongly influence by the helicity of the initial conditions or the system forcing. Our results also provide an important link between quantised vortex reconnections and reconnections in classical and magnetised fluids.
\end{abstract}

\pacs{ 47.32.C- (Vortex dynamics) 47.32.cf (Vortex reconnection and rings), 47.37.+q (Hydrodynamic aspects of superfluidity)}
\maketitle
Reconnection is a term used in both the hydro- and magnetohydrodynamic communities, when referring to topological changes in the system due to dissipation through either viscosity or resistivity respectively \cite{Kida1994,priest2007magnetic}. 
The importance of vortex reconnections in turbulence 
cannot be understated.  In hydrodynamics vortex reconnections play a role in the
energy cascade, contribute to the fine-scale mixing and are the
dominant mechanism of jet noise generation \cite{Hussain83,Hussain86}. Magnetic reconnections are believed to be crucial in understanding the anomalous heating of the solar corona \cite{priest2007magnetic}.
An important measure of the topology of either the vorticity or magnetic field is the fluid or magnetic helicity \cite{MoffattHelicity}; in 3D both quantities are conserved in the ideal systems (where reconnections are prohibited). In the limit of small viscosity or resistivity, typical of many natural systems, it is perhaps reasonable to assume that helicity is approximately conserved in reconnection events. If so natural questions to ask are how the dynamics are constrained by helicity conservation, and how is helicity transformed by reconnections. For example conservation of helicity is a crucial aspect of the relaxation of a plasma to a Taylor state. In general however observations or experiments which are able to probe the topology of vortices in classical hydrodynamics or magnetic structures in MHD are extremely challenging. 

In this letter we turn our attention to the reconnection of quantised vortices, found in superfluid helium and atomic Bose-Einstien condensates (BECs) \cite{barenghi2001quantized}. These are particularly attractive systems as vortices are stable topological defects, with a fixed size and circulation, due to the constraints of quantum mechanics. There is a long history of the study of reconnections in quantum fluids, including analytic  \cite{West2003}, numerical \cite{Koplik} and experimental studies \cite{Bewley}. However it is surprising that the question of helicity transfer or conservation has not been considered. In Particular atomic BECs provide a perfect setting to experimentally study vortex reconnection. However if helicity is destroyed during a reconnection, then it is difficult to see how any deep comparisons could be drawn with reconnections between classical vortices or magnetic reconnection; the system would perhaps be \emph{too} simple. 

Before proceeding further we briefly review some of the properties of quantised vortices. The governing equation for a weakly interacting BEC is the Gross-Pitaevski equation (GPE)

 \begin{equation}\label{eq:GPE}
 2i \dfrac{\partial \psi}{\partial t}=-\Lapl \psi +\left ( |\psi|^2-1\right )\psi,
 \end{equation}
 written here in dimensionless form and in the absence of an external trapping potential. If we apply the Madeleung transformation $\psi = \sqrt\rho \rme^{\rmi \theta}$ then we obtain the familiar continuity equation for a compressible fluid, and the following modified Euler equation

\begin{equation}\label{eq:GPEinNSmom}
\rho \left(\Dpar{u_i}{t}+u_j\Dpar{u_i}{u_j} \right) =
-\Dpar{p}{x_i} + \Dpar{\tau_{ij}}{x_j},\quad\quad i=1,2,3.
\end{equation}
Here $\rho=|\psi|^2$, $\bu=\Grad{\theta}$, $p=\rho^2/4$ and 
\begin{equation}
\tau_{ij}=\Fra \rho 4 \Dparxy{\ln\rho}{x_i}{x_j}
\end{equation}
is the so-called quantum stress; it is this term which is responsible for vortex reconnections. Without this term (which is negligible compared to the pressure term at length scales larger than unity, the coherence length) the helicity would be conserved. It acts to regularise the momentum equation, preventing singularities which may arise in an ideal (Euler) fluid.

\begin{figure}[h!!!]
\begin{center}
\includegraphics[width=0.4\textwidth]{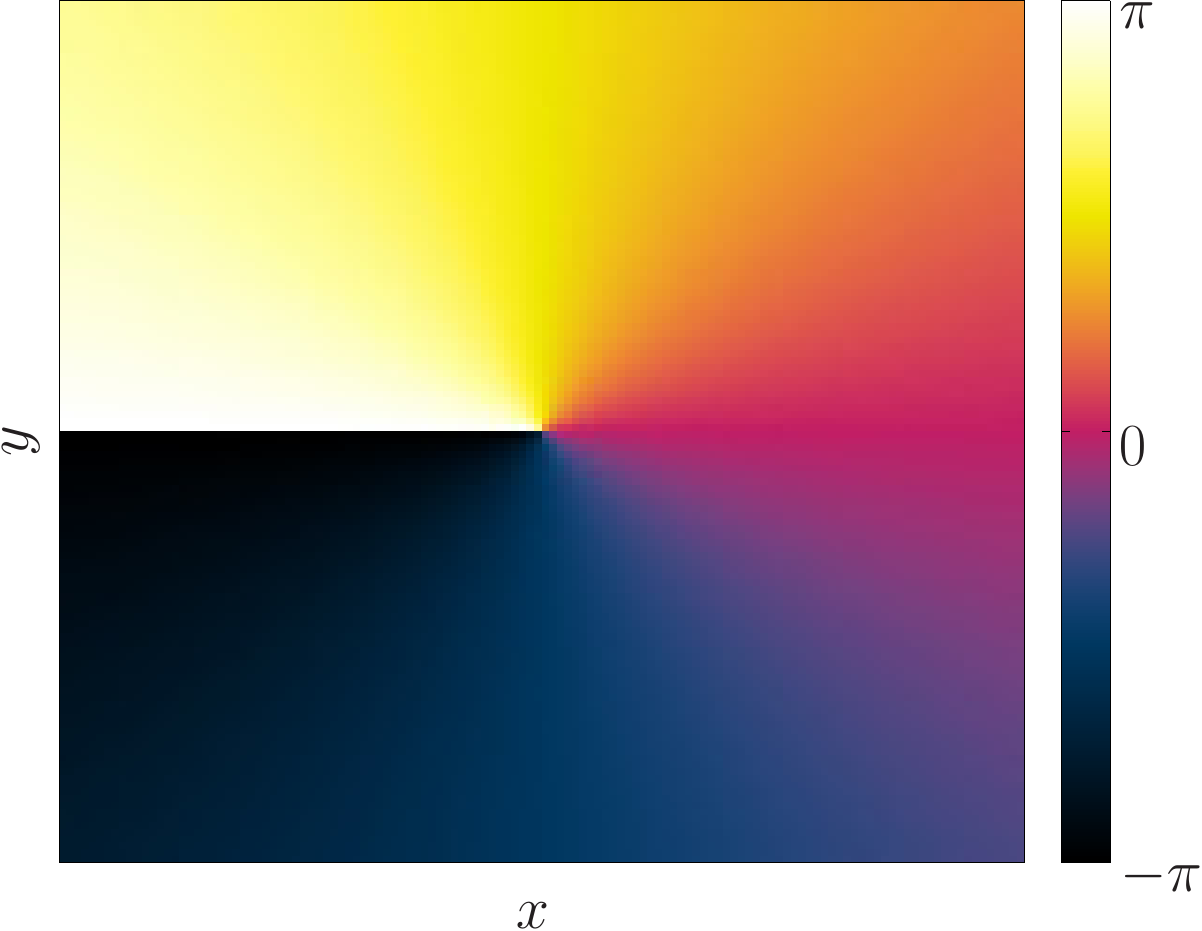}
\end{center}
\caption{(Color online) The $2\pi$ winding of the phase, $\theta$ around an isolated straight vortex.}
\label{fig:1}
\end{figure}

Rotational motion in a quantum fluid is confined to topological defects, vortex lines where the density is zero and around which the circulation is fixed by the condition

\begin{equation}
\oint_C \Vmy{u} \cdot \Vmy{dr}=\frac{h}{m}=\kappa,
\label{eq:circulation}
\end{equation}

\noindent
where $C$ is a closed integration path around the vortex axis, $h$ is Planck's
constant,  $m$ is the mass of the relevant boson and $\kappa$ is the quantum of circulation.
The circulation is associated with a $2\pi$ winding of the phase, $\theta$, see Fig.~\ref{fig:1}.
In principal multiply-charged quantised vortices can exist (i.e.~with circulation $n\kappa$, $n>1$) however such vortices are unstable and breakup into singly-charged vortices \cite{Law2008}.

For a simple example, which provides an insight into the importance of helicity conservation consider the situation of two linked magnetic flux tubes in a highly conducing plasma. Due to the linkage the magnetic helicity is non-zero. Magnetic tension will cause the rings to contract until they reconnect. The link will be broken and if we assume the majority of the magnetic helicity is conserved it must be transformed into both writhe and twist. This argument follows from Cal\u{u}garean\u{u} \cite{Moffatt09111992} who showed that the linking number of two closed ribbons can be decomposed as the sum of writhe and twist (where here twist is the sum of the internal twist of the ribbon and the torsion). The internal twist is a twisting of the magnetic field lines within the flux tube and is possible due to the internal structure of the flux tubes. Let us now consider two linked quantised vortices. Their self induced motion will drive a reconnection and again the link is broken. However it seems that as the vortices a simply lines of zero density their is no internal structure which can become twisted. In this letter we shall argue that there is an internal structure to the vortices and show that the reconnection of linked quantised vortices results in a `twisting' of the vortices.

\begin{figure}
\begin{center}
\includegraphics[width=0.3\textwidth]{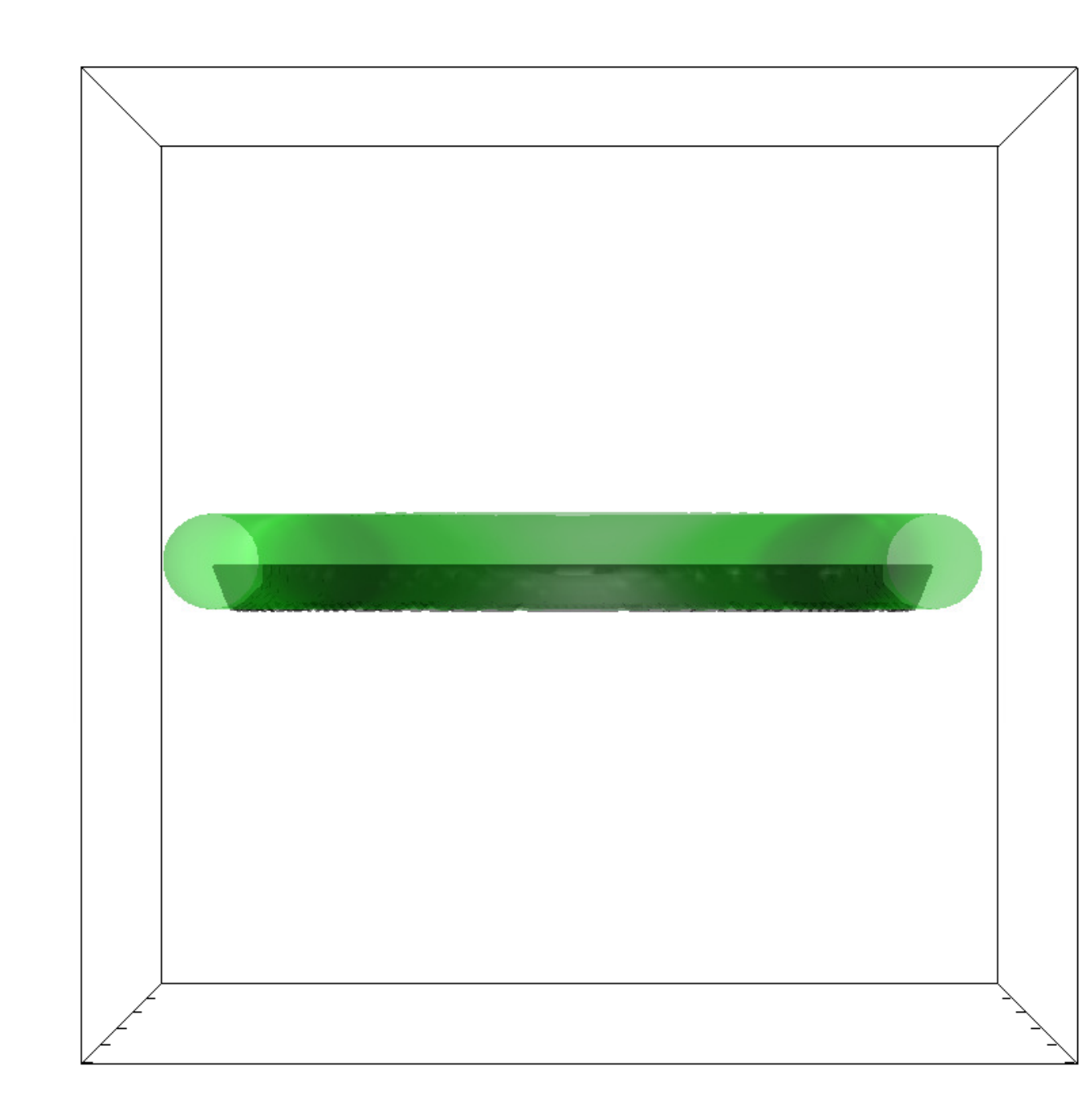}
\includegraphics[width=0.125\textwidth]{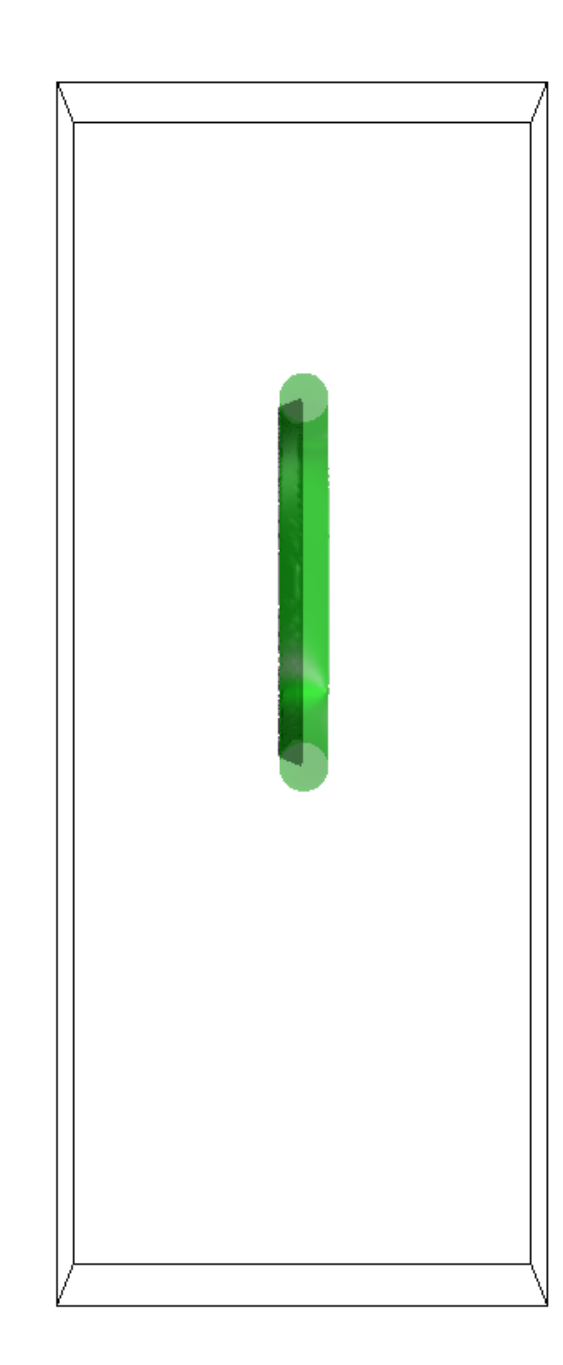}
\end{center}
\caption{(Color online) An isosurface displaying the low density regions in a BEC with a single quantised vortex ring in the $xy$-plane, viewed in the $xy$-plane (left) and $xz$-plane (right). The black curve denotes a line of constant phase.}
\label{fig:2}
\end{figure}
To this end we perform a numerical simulation of Eqn.(\ref{eq:GPE}). Space is disctretized onto a $N=256^3$ uniform cartesian mesh with $x_i \in [-32,32]$, $i=1,2,3$. Spacial derivatives are approximated by a $6^{\rm th}$--order finite difference scheme and a $3^{\rm rd}$--order Runge-Kutta scheme is used for time evolution. Reflective boundaries are imposed in all three cartesian directions. 
A single vortex ring in the $xy$-plane at the origin with radius $R_0$ can be initialised by taking the following form for the wavefunction $\psi$ at $t=0$,

\begin{figure*}
\begin{center}
\includegraphics[width=0.4\textwidth]{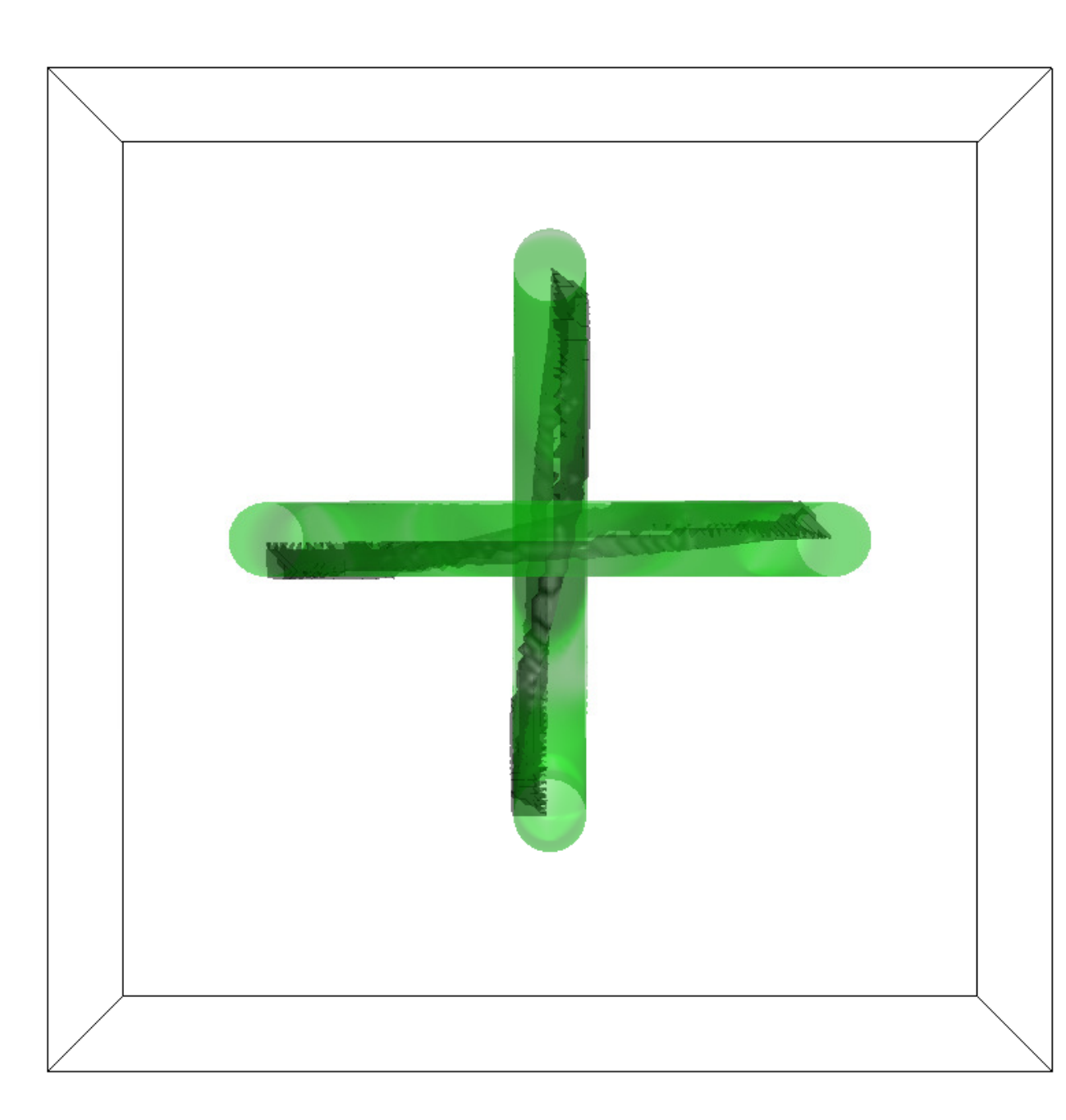}
\includegraphics[width=0.2\textwidth]{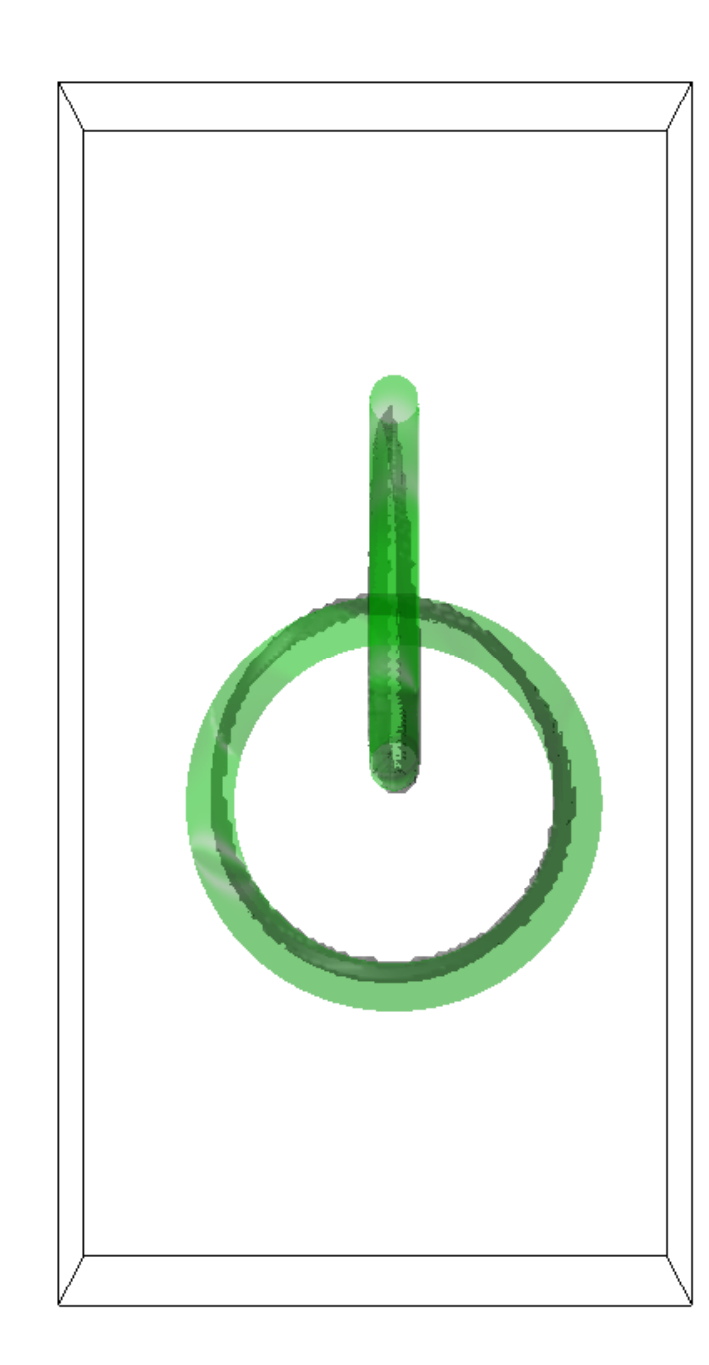}
\includegraphics[width=0.2\textwidth]{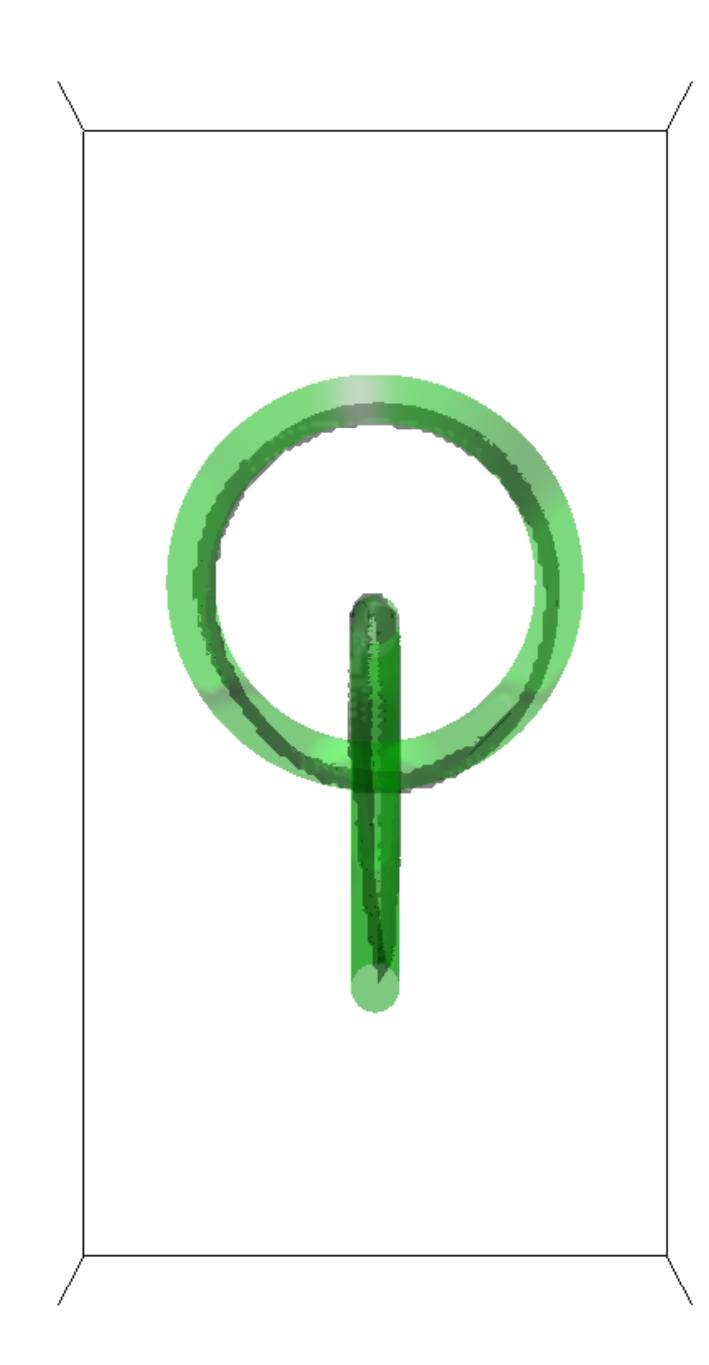}
\end{center}
\caption{(Color online) An isosurface displaying the low density regions in a BEC with a two linked quantised vortex rings, viewed in the $xy$-plane (left), $yz$-plane (centre) and $xz$-plane (right). The black curve denotes a line of constant phase, which displays non-trivial features when compared to the case of a single vortex ring (Fig. \ref{fig:2}, although there is no net twisting of either vortex ring. The box is purely for the purposes of visualisation.}
\label{fig:3}
\end{figure*}

\begin{figure}
\begin{center}
\includegraphics[width=0.35\textwidth]{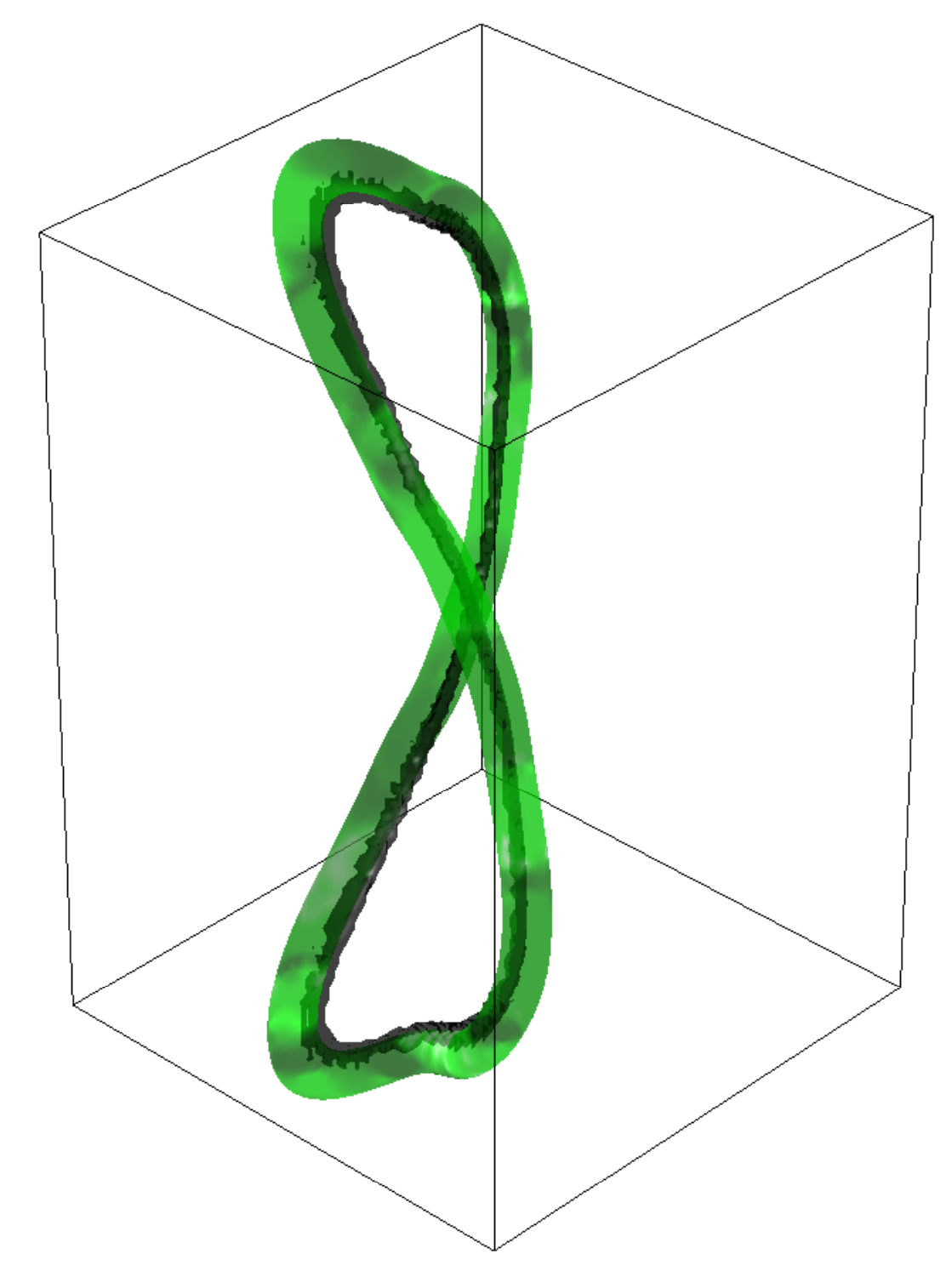}\\
\includegraphics[width=0.4\textwidth]{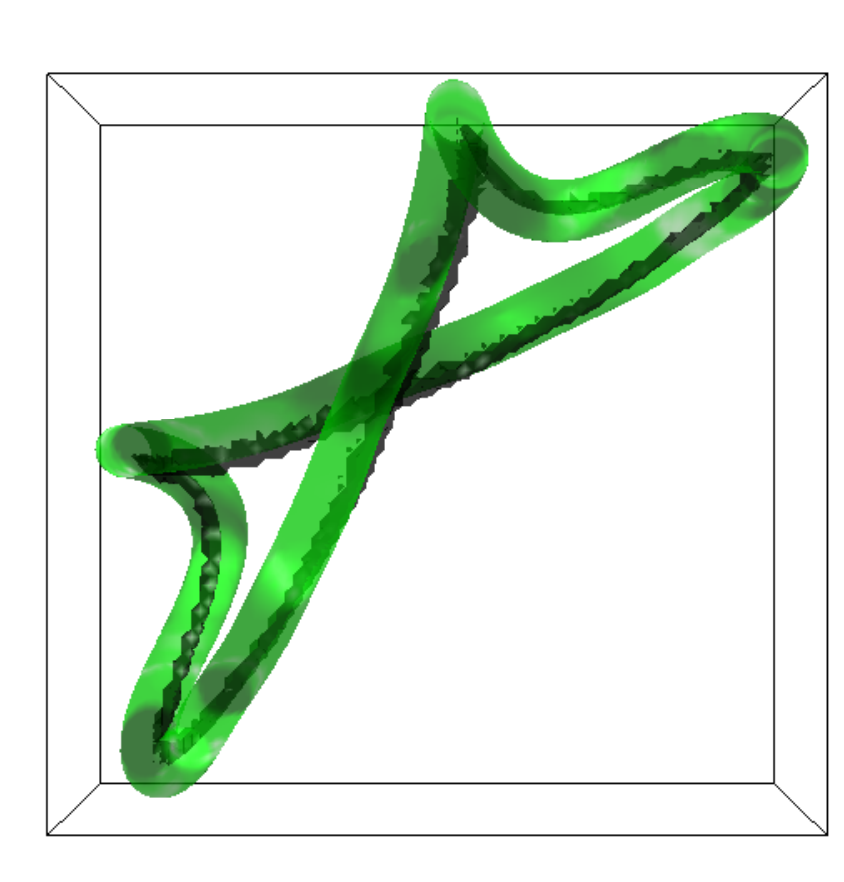}
\end{center}
\caption{(Color online) An isosurface displaying the low density regions after the reconnection of two linked quantised vortices. The result is a single vortex ring with Kelvin wave perturbations. The black curve denotes a line of constant phase, which can be seen to twist along the vortex.}
\label{fig:4}
\end{figure}

\begin{figure}
\begin{center}
\includegraphics[width=0.45\textwidth]{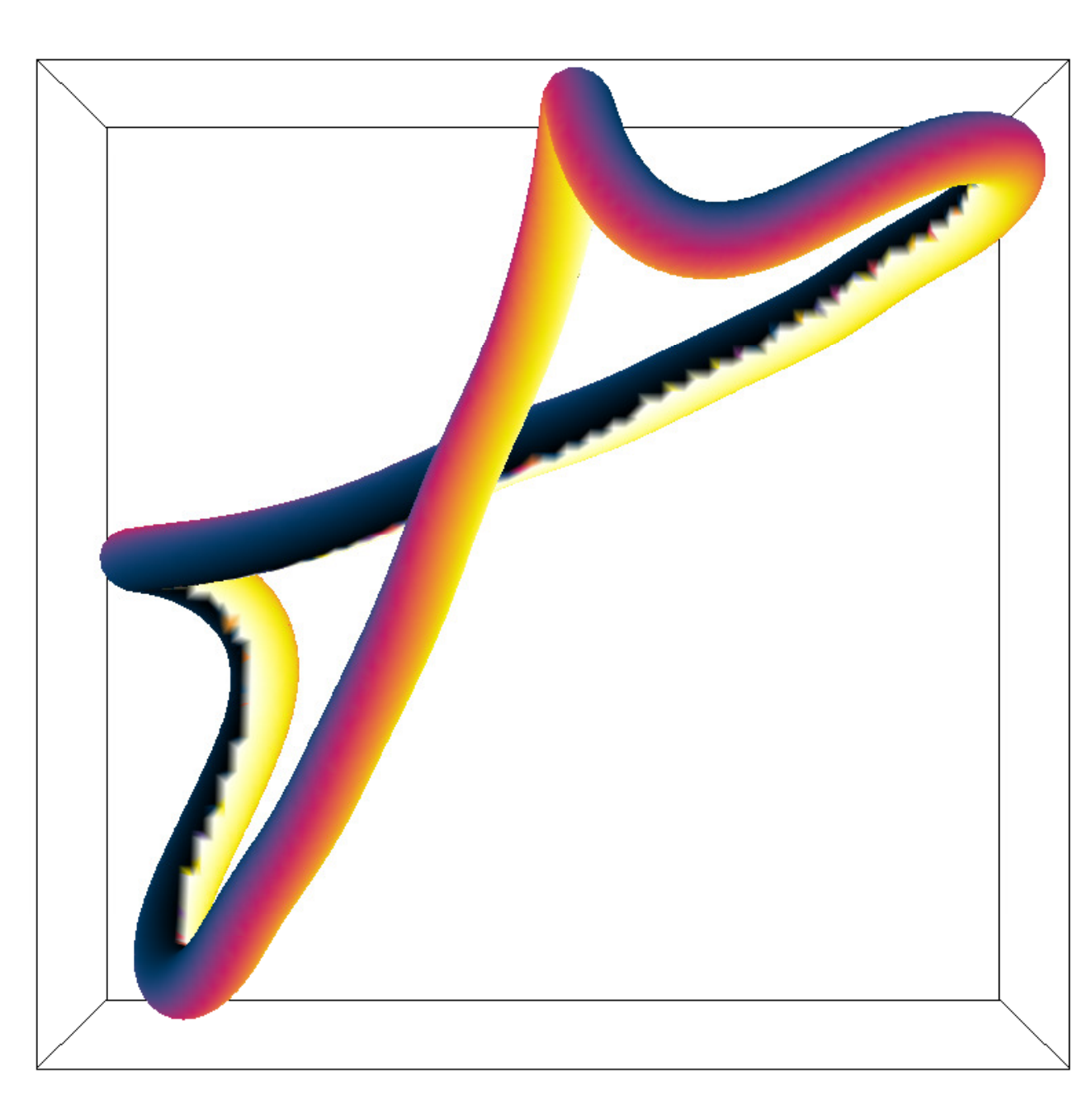}
\end{center}
\caption{(Color online) The resulting quantised vortex ring, as presented in Fig. \ref{fig:4}, but now coloured according the the local phase (see Fig.~\ref{fig:1}).}
\label{fig:5}
\end{figure}

 \begin{equation}
 \psi(\bx)=\Psi(z,s+R_0)\Psi^*(z,s-R_0), 
  \end{equation}
 \begin{equation}
  \Psi(z,s)=f\left( \sqrt{z^2+s^2}\right ) \exp(i\theta).
 \end{equation}
 
Here $f(r)=r(c_1+c_2r^2)^{1/2}/(1+c_3r^2+c_2r^4)^{1/2}$, and $s=\sqrt{x^2+y^2}$. The constants $c_i$ are found by seeking a Pad\'{e} approximation to the density profile of an isolated vortex \cite{berloff_pade}; $c_1=11/32$,
$c_3=(5-32 c_1)/(48-192 c_1)$, $c_2=c_1\left( c_3-1/4 \right)$. Figure \ref{fig:2} shows a single vortex ring, with a line of approximately constant phase. The vortex ring is visualised by plotting an isosurface of the lowest density regions ($|\psi|^2<0.4$). In order to visualise the orientation of the phase we compute the following field $Z=\theta \times H(0.6-|\psi|^2)$, where $H$ is the Heaviside function. This leaves purely the phase inside the regions of lowest density, the value 0.6 is chosen purely to optimise the visualisation of the vortex structure. We overlay an isosurface of $Z$ at values larger than $0.8\pi$ ($\theta \in [-\pi, \pi]$) to the isosurface of the condensate density; this approach simply overlays a line of approximately constant phase on the visualisation of the vortex rings. From this visualisation it is clear that a single vortex ring with the initialisation above does not show any twisting of the phase plane as we move along the ring.

The extension of this initial condition to rings in other planes, translations of the rings and multiple rings is straightforward. We initialised two linked vortex rings of radius $R_0=10$ centred at $x=y=0$, $z=0$ \& $12$, the lower ring is in the $yz$-plane, whilst the upper ring is in the $xz$-plane. Figure \ref{fig:3} shows a visualisation of the initial vortex rings again with a indication of the rotation of the phase plane along the vortex. In contrast to a single vortex ring the phase is non-trivial, and whilst there is no rotation of the phase (i.e. the only contribution to the helicity is the linking of the rings) the kink in the phase is essential to allow a transfer of helicity upon reconnection.

As the simulation progresses the self induced motion of the vortex rings drives a reconnection. During the reconnection we observe the generation of a pressure (density) rarefaction wave which has been previously reported in the literature \cite{zuccher_recon}. As a result of the reconnection we are left with a single vortex ring with noticeable helical perturbations, Kelvin waves. Of course the important question for the purpose of this study is the reconnections effect on the phase. An indication of the generation of internal twist can be seen in the phase of the two rings before reconnection, as shown in Fig.\ref{fig:3}. If one was to go in `by hand' and connect the two black curves of constant phase, then the resulting configuration would indeed be twisted.  

Figure \ref{fig:4} shows the visualisation (described above) of the single vortex ring after reconnection. A visible twisting of the curve of constant phase around the vortex is clear in the two projections presented. To further visualise this twisting of the phase in Fig. \ref{fig:5} we plot an isosurface of the vortex ring coloured by the phase. Our conclusions are thus clear and simple, quantised vortex reconnection is capable of transforming helicity from one `form' to another. In particular the reconnection of linked vortices leads to a twisting of the resulting vortices, which manifests itself as a rotation of the phase plane along the vortex.

We now turn our attention to the consequences of this result. As discussed earlier the velocity is a gradient of the phase,  $\bu=\Grad{\theta}$, and hence twisting of the phase along the ring will give rise to an axial flow along the vortex ring. Of course this will have an important effect on the dynamics of the system, for example such a background flow will alter the velocity of Kelvin waves propagating along the vortex. Kelvin waves are crucial in the decay of quantum turbulence however as the possibility of twist of the vortices has not been discussed previously it has not been accounted for in any of the Kelvin wave theories \cite{BaggaleyKWC}.  In addition the flow which a twisted vortex induces will be helical and so the nature of a flow induced by a tangle of twisted vortices will be very different from the untwisted case. Hence the results presented here have important consequences for the dynamics of quantum turbulence, which is a burgeoning field in itself \cite{VinenNiemela}. For example one would envisage that the evolution of helical and non-helical systems differs dramatically, as has been shown in MHD simulations \cite{DelSordo2010,Candelaresi2011}. One would also imagine that it would be fruitful to investigate topological constraints on the evolution of a tangle of quantised vortices. In addition twist helicity could lead to instabilities of the vortex which have not yet been investigated in the literature.

However our results are of wider interest. As we remarked at the start of this letter BECs provide an ideal system for the experimental study vortex reconnection. The results presented here suggest that the system is capable of exhibiting the topological complexity of classical and magnetised fluids, which is clearly important if any knowledge gained from experiments with BECs is to be applied to these two systems.
In a more general setting twist and writhe provide a universal language to understand complex topologies of tubes or ribbons and are commonly applied to DNA and proteins \cite{Fuller1971}. Until now no direct link to quantum fluids could be made.
Finally we remark that experimental techniques for measuring the phase of a BEC exist \cite{Denschlag2000}, and we hope that the results presented here can be confirmed experimentally in the future.

\begin{acknowledgments}
We acknowledge fruitful discussions with Carlo Barenghi, Sergey Nazarenko,  Jason Laurie and Hayder Salman.
\end{acknowledgments}

\bibliographystyle{pf}
\bibliography{biblio}
\end{document}